\documentclass{raa}           

\usepackage{graphicx,times}
\usepackage{natbib}
\usepackage{amssymb,amsmath}
\bibpunct{(}{)}{;}{a}{}{,}

\usepackage[pagebackref=true]{hyperref}

\begin{document}

\title{Chemical tagging of tidal tail star candidates of NGC~6362}

\volnopage{ {\bf 20XX} Vol.\ {\bf X} No. {\bf XX}, 000--000}
\setcounter{page}{1}

\author{Andr\'es E. Piatti\inst{1}}

\institute{ Instituto Interdisciplinario de Ciencias B\'asicas (ICB), CONICET-UNCuyo, Padre J. Contreras 1300, M5502JMA, Mendoza, Argentina; {\it andres.piatti@fcen.uncu.edu.ar}\\
\and Consejo Nacional de Investigaciones Cient\'{\i}ficas y T\'ecnicas (CONICET), Godoy Cruz 2290, C1425FQB,  Buenos Aires, Argentina\\
\vs \no
{\small Received 20XX Month Day; accepted 20XX Month Day}
}

\abstract{The inner Milky Way disk globular cluster NGC~6362 appears to exhibit
tidal tails  composed of stars that have  proper motions and positions in the color-magnitude diagram similar to those of cluster stars. Because recent results seem  also to show that these
stars are distributed across the regions least affected by interstellar absorption and
reproduce the observed composite star field density map, we
carried out a detailed spectroscopic analysis of a number of chemical element abundances of
tidal tail star candidates in order to investigate the relationship of them with NGC~6362. 
From European Southern Observatory's VLT@FLAMES spectra we found 
that the red giant branch stars selected as cluster's tidal tail stars do not have
overall metallicities nor abundances of Mg, Ca, Sc, Ti, Cr, Ni and Ba similar to the cluster's ones.
Moreover, they are mainly alike to stars that belong to the Milky Way thick disk, some of them 
could be part of the thin disk and a minor percentage could belong to the Milky Way halo star 
population. On the other hand, since the resulting radial velocities do not exhibit a distribution 
function similar to that of cluster's stars, we concluded that looking for kinematic properties similar 
to those of the cluster would not seem to be an approach for selecting cluster's tidal tail stars
as suitable as previously thought.
}

\authorrunning{Andr\'es E. Piatti}            
\titlerunning{Chemical tagging of tidal tail stars}  
\maketitle

%
\section{Introduction}           

Recently, \citet{zhangetal2022} published a stringent compilation of Milky Way globular clusters 
with robust  detection of extra-tidal structures. Their catalog includes 46 globular clusters
classified as follows: 27 with tidal tails, 4 with extended envelopes, and 15 without observed 
extended features.
The detection of tidal tails around Milky Way globular clusters is a research avenue of significant importance
for our understanding of a wide variety of issues. For instance, globular clusters 
associated to destroyed dwarf progenitors should show tidal tails 
\citep{carballobelloetal2014,mackeyetal2019}; globular clusters formed in dark matter 
mini-halos should present tidal tails with a relatively large velocity dispersion 
\citep{malhanetal2021}; the extension and shape of tidal tails tell us about the
dynamical history of a globular cluster as a consequence of its interaction with the Milky Way 
\citep{hb2015,deboeretal2019}, etc. 

The innermost globular cluster in the \citet{zhangetal2022}'s compilation is NGC~6362 
\citep[galactocentric distance $R_{GC}$ = 5.5 kpc;][]{bv2021}, which shows 
tidal tails.  \citet{zhangetal2022} included NGC~6362 as a globular cluster with tidal tails 
based on the work of \citet{kunduetal2019}, who found 73 highest-ranked extra-tidal red giants
with {\it Gaia} DR2 proper motions within 3-sigma around the mean cluster proper motion.
However, escaping stars should show a dispersion velocity larger than that for outermost
bound stars \citep{wanetal2021,piatti2023b}, which is not the case of these 73 extra-tidal 
giant candidates. On 
the other hand, NGC~6362 should not have tidal tails composed by kinematically 
cold stars, because  \citet{kunduetal2019} showed that the orbit of NGC~6362 is chaotic, which
means that it washes out (accelerating) those stars \citep{mestreetal2020}.
What we mention above is a strong motivation for deciphering whether NGC~6362 already
pertain to the group of globular clusters with observed tidal tails, making it a compelling 
science case because its position in the inner Milky Way. 
Precise abundances of some chemical elements for \citet{kunduetal2019}'s highest-ranked 
extra-tidal stars could provide the final link between the extra tidal star candidates 
and NGC~6362. Indeed, we assumed that chemical abundances significantly deviating from the 
mean values known for NGC~6362 point to a star formation scenario not related to that of
the cluster itself.

Precisely, we embarked in an observational campaign with the aim of obtaining spectra
of these selected stars with the aim of deriving abundances of different chemical elements
and hence to investigate their relationship with NGC~6362. Likewise, the results will
be helpful to assess on the validity of some criteria used to select these stars as
highest-ranked extra-tidal stars. In other words, we attempt to shed light about 
the suitability of the frequently applied 
criterion to detect  tidal tails based on similar proper motions for cluster 
members and tidal tail stars \citep[e.g.,][]{sollima2020}. In Section~2 we present the
acquired observational material, and in Section~3 we derive different astrophysical
properties of the target stars. Section~4 is devoted to the discussion on their chemical element
abundances. Finally, Section~5 summarizes the main conclusions of this work.

\begin{figure*}
\includegraphics[width=\textwidth]{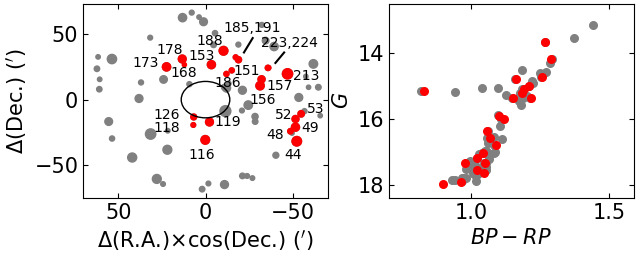}
\caption{Spatial distribution of \citet{kunduetal2019}'s tidal tail stars (left panel), and 
their location in the color-magnitude diagram (right panel). Red symbols represent the stars
observed in this work. In the left panel, the size of the symbol is proportional to the star's 
brightness, and the black circle represents the cluster's Jacobi radius \citep{morenoetal2014}.}
\label{fig1}
\end{figure*}

\section{Observational data}

The red giant stars selected by \citet{kunduetal2019} as tidal tail candidates of 
NGC~6362 have proper motions within 3$\sigma$ of the mean cluster
proper motion \citep{vasiliev2019}. They are also located beyond the cluster Jacobi radius 
\citep{morenoetal2014}, and are distributed in the color-magnitude diagram
along the cluster red giant branch. Figure~\ref{fig1} illustrates their spacial distribution
and their location in the cluster color-magnitude diagram \citep[figure built from Table~3 of][]{kunduetal2019}.
We accommodated as many stars as possible across the FLAMES spectrograph's
field-of-view \citep{pasquinietal2002} attached at the European Southern Observatory (ESO)'s 
Very Large Telescope (programme 113.2661.001, PI:Piatti). By applying the completeness 
expression of \citet{p17a}, which allowed us to take into account the tidal tail stars' 
spatial coverage and their distribution in the color-magnitude diagram, we found that the 
finally selected star sample, represented with red circles in Figure~\ref{fig1}, 
statistically represents $\sim$ 80$\%$  of the whole \citet{kunduetal2019}'s red giant 
sample brighter than $G$ = 18.0 mag.

We used the FLAMES@GIRAFFE spectrograph with the MEDUSA 132 fiber component and the HR14 grism 
($\sim$ 6290-6690 \AA), which gives a dispersion of 0.05 $\AA$/pixel with a resolution of 
R = 17000, appropriate to measure abundances of different chemical elements with an accuracy 
enough to chemically distinguish stars formed in NGC~6362 or elsewhere. 
We obtained spectra of the selected stars from a total exposure time of 2235 sec per 
observed science field (five different pointings), and typical airmass and seeing of 
$\sim$1.34-1.45 and 0.45$\arcsec$-0.76$\arcsec$, respectively. Fifteen sky spectra were
also taken simultaneously to each science field.

The data were processed in the standard way, following the FLAMES@GIRAFFE ESO 
pipeline\footnote{http://www.eso.org/sci/software/pipelines/}, which includes zero-subtraction, 
flat-field correction, wavelength calibration with a standard Th-Ar lamp, extraction of
one-dimensional spectra, etc.  By comparing the observed position of several sky emission lines 
with their rest-frame position using the sky lines atlas by \citet{oetal1996}, we checked that
there does not exist any significant wavelength shifts. Finally, the sky-subtracted spectra were
extracted from all the multi-object spectroscopy exposures. Table~\ref{tab1} lists the observed
stars alongside with their equatorial coordinates, {\it Gaia} DR3 \citep{gaiaetal2016,gaiaetal2022b} 
photometry, and average signal-to-noise ratio (S/N)
along the whole wavelength range of each spectrum. One science field, which includes stars $\#$116,
118, 119 and 126 was observed in two different night, the first one under no good weather conditions.
From this night, only the spectrum of $\#$116 could be extracted with a reasonable S/N ratio.
Therefore, we used the spectra of star $\#$116 coming from both nights to check internal consistency 
in the radial velocity and chemical element abundance measurements.

Finally, we retrieved from the ESO archive\footnote{https://archive.eso.org/eso/eso\_archive\_main.html}
the processed spectra of NGC~6362's red giants taken with FLAMES@GIRAFFE and an instrument setup similar 
to ours  (programme 093.D-0618(A), PI: Dalessandro). The main aim of dealing with this data is to validate 
our procedure of deriving abundances of chemical elements, by comparing the present resulting values with 
those of \citet{mucciarellietal2016}

\begin{table}
\caption{Astrophysical properties of studied stars.}
\label{tab1}
\setlength{\tabcolsep}{1pt}
\small
\begin{tabular}{rccccccrcccr}
   \hline\noalign{\smallskip}

ID &     R.A.         &    Dec.            &  $G$   & $BP$  & $RP$  & $<$S/N$>$ &RV\hspace{0.5cm}     & $T_{eff}$ & log $g$  & $v_t$  & [Fe/H]\hspace{0.3cm} \\
   & ($\degr$)        & ($\degr$)          & (mag)  & (mag) & (mag) &           &(km/s)\hspace{0.3cm} & 
(K) &          & (km/s) & (dex)\hspace{0.3cm}  \\
\hline\noalign{\smallskip}
44 & 260.696624999996 & -67.57813888888806 & 14.16  & 14.72 & 13.45 & 65.8      & -24.67$\pm$0.37     &  4800     & 2.80     & 1.30   &  0.00$\pm$0.05 \\
48 & 260.865999999996 & -67.45077777777699 & 17.03  & 17.46 & 16.44 & 19.2      & -22.93$\pm$0.46     &  5250     & 3.20     & 2.00   &  0.00$\pm$0.08 \\
49 & 260.746333333329 & -67.39899999999916 & 15.20  & 15.70 & 14.53 & 34.1      &-100.03$\pm$1.18     &  4800     & 1.50     & 1.55   & -1.37$\pm$0.08 \\
52 & 260.755333333329 & -67.29455555555468 & 16.00  & 16.46 & 15.36 & 30.7      & 208.71$\pm$0.84     &  5050     & 2.60     & 4.00   & -1.13$\pm$0.06 \\
53 & 260.621541666662 & -67.22872222222135 & 16.36  & 16.79 & 15.75 & 24.4      & -22.03$\pm$1.15     &  5250     & 3.00     & 2.65   & -1.30$\pm$0.15 \\
116& 262.984874999996 & -67.56041666666580 & 14.78  & 15.26 & 14.12 & 56.3      & -92.06$\pm$0.41     &  4870     & 2.80     & 1.50   & -0.43$\pm$0.06 \\
118& 263.281416666662 & -67.37069444444360 & 17.63  & 18.07 & 17.04 & 19.4      &  -8.83$\pm$0.73     &  5180     & 3.80     & 1.50   & -0.42$\pm$0.09 \\
119& 262.881083333329 & -67.33286111111025 & 15.33  & 15.85 & 14.60 & 43.3      & -13.82$\pm$0.76     &  4880     & 2.40     & 1.00   & -0.66$\pm$0.09 \\
126& 263.269999999996 & -67.26908333333249 & 16.76  & 17.22 & 16.15 & 25.2      & -13.75$\pm$0.99     &  5100     & 3.40     & 2.00   & -0.70$\pm$0.10 \\
151& 262.474458333329 & -66.72133333333246 & 17.33  & 17.73 & 16.77 & 23.8      &  50.12$\pm$1.34     &  5300     & 1.80     & 1.50   &  0.02$\pm$0.13 \\
153& 262.346874999996 & -66.67613888888799 & 17.33  & 17.76 & 16.72 & 23.9      &  -3.34$\pm$0.83     &  5080     & 3.50     & 1.20   & -0.81$\pm$0.06 \\ 
156& 261.654583333329 & -66.87124999999916 & 14.99  & 15.50 & 14.31 & 67.0      &  46.16$\pm$0.48     &  4860     & 2.40     & 1.50   & -0.33$\pm$0.07 \\ 
157& 261.622874999996 & -66.78972222222137 & 15.88  & 16.32 & 15.23 & 43.7      &-111.37$\pm$0.64     &  5170     & 3.10     & 1.00   & -0.20$\pm$0.09 \\
168& 263.486999999996 & -66.60394444444357 & 17.96  & 18.30 & 17.43 & 16.5      & -70.81$\pm$1.00     &  5550     & 3.30     & 4.00   & -0.55$\pm$0.05 \\ 
173& 263.916249999996 & -66.63174999999915 & 15.06  & 15.56 & 14.39 & 67.2      & -81.34$\pm$0.57     &  4930     & 2.50     & 1.20   & -0.67$\pm$0.05 \\
178& 263.536291666662 & -66.53094444444356 & 15.33  & 15.81 & 14.68 & 39.3      &  39.50$\pm$0.44     &  4900     & 2.60     & 1.20   & -0.90$\pm$0.06 \\
185& 262.257874999996 & -66.50902777777696 & 17.53  & 17.96 & 16.96 & 23.0      &-106.39$\pm$1.34     &  5230     & 3.60     & 5.00   & -0.86$\pm$0.08 \\
186& 262.837666666662 & -66.60333333333248 & 15.15  & 15.46 & 14.65 & 50.8      &  34.48$\pm$0.50     &  5710     & 3.10     & 5.00   & -1.00$\pm$0.15 \\ 
188& 262.549666666662 & -66.42669444444357 & 14.71  & 15.25 & 14.01 & 65.2      &  15.91$\pm$0.39     &  4850     & 2.20     & 1.25   & -0.20$\pm$0.07 \\
191& 262.189499999996 & -66.54047222222139 & 16.56  & 17.00 & 15.95 & 33.4      &  34.45$\pm$0.63     &  5140     & 3.15     & 5.00   & -0.49$\pm$0.12 \\ 
213& 260.999708333329 & -66.71997222222140 & 13.65  & 14.20 & 12.95 & 88.3      & -71.84$\pm$0.30     &  4820     & 2.50     & 1.40   & -0.16$\pm$0.08 \\
223& 261.477583333329 & -66.64430555555467 & 17.17  & 17.59 & 16.59 & 22.8      &  56.95$\pm$0.87     &  5330     & 3.80     & 2.40   & -0.05$\pm$0.06 \\
224& 261.457291666662 & -66.64280555555466 & 17.90  & 18.29 & 17.35 & 18.6      & -27.65$\pm$0.73     &  5360     & 3.60     & 5.00   & -0.60$\pm$0.10 \\
\noalign{\smallskip}\hline

\end{tabular}
\end{table}

\section{Stellar parameters and abundance analysis}
\subsection{Radial velocity measurements}

We measured radial velocities (RVs) by cross-correlating the observed spectra and 
the Arcturus spectrum taken from 
\citet{hinkleetal2000}\footnote{https://www.eso.org/sci/observing/tools/uvespop/bright$\_$stars$\_$uptonow.html}.
All the spectra were continuum normalized before the cross-correlation procedure and 
the Arcturus spectrum spectral resolution was degraded to match the resolution of our
science spectra. We made use of the  IRAF.FXCOR task, which implements the algorithm described 
in \citet{tonry1979} for the construction of the cross-correlation function of each (object, Arcturus) 
spectra pair. In addition to the RVs,  FXCOR returns the CCF normalized peak ($h$), which is an 
indicator of the degree of similarity between the correlated spectra and the Tonry \& Davis ratio (TDR) 
defined as TDR$=h/(\sqrt{2}\sigma_{a})$, where $\sigma_{a}$ is root mean square of the CCF antisymmetric component. 
The resulting RVs are associated to $h$ values greater than 0.8.  We finally carried 
out the respective heliocentric corrections by using the IRAF task  RVCORRECT. 
Table~\ref{tab1} lists the resulting heliocentric RVs with their respective uncertainties. 
The difference in the RVs measured for star $\#$116, observed in two different nights, resulted to 
be 0.00$\pm$0.41 km/s, where the error comes from the propagation of both individual RV uncertainties. 
We compared our resulting RVs with those of {\it Gaia} DR3 
for four stars in common ($\#$ 44, 116, 188 and 213)) and found a difference ({\it Gaia} DR3 - our) 
of 0.14$\pm$1.28 km/s.

\subsection{Stellar atmospheric parameters}

With the aim of obtaining initial estimates of effective temperature ($T_{eff}$) and
surface gravity (log $g$) for chemical abundance analysis, we
derived photometric parameters using the {\it Gaia} DR3 photometry (see Table~\ref{tab1}).
The effective temperatures were computed from {\it Gaia} $BP-RP$ colors, previously
corrected by interstellar absorption using the $E(B-V)$ values of \citet{piatti2024}
and the total to selective absorption ratios given by \citet{chenetal2019}. The
reddening corrected ($BP-RP$)$_o$ colors were then transformed to Johnson ($R-I$)$_o$ ones
using the transformation equations derived by \citet{pancinoetal2022}. Finally, we employed
the suitable correlation between the free metallicity sensitivity
Johnson ($R-I$)$_o$  color and $T_{eff}$ derived by \citet{alonsoetal1999} 
The surface gravity was calculated using the equation log $g$ = 4.44 + log $M_*$ +
4log $T_{eff}$/5780 + 0.4($V_o$ - $(m-M)_o$ + $BC(V)$ - 4.75) \citep{vennetal2017}, 
where $M_*$ = 0.75 is the typical mass of an old red giant branch star; $V_o$ is the dereddened
Johnson $V$ magnitude (calculated from \citep{pancinoetal2022}'s transformation equations); 
$(m-M)_o$ = 14.42 mag 
\citep{bv2021} is the true cluster distance modulus; and $BC(V)$ is the 
\citet[][eq. 18]{alonsoetal1999} bolometric correction, for which we adopted a cluster mean 
metallicity [Fe/H] = -1.07$\pm$0.01 dex \citep{massarietal2017}.


We derived spectroscopic $T_{eff}$, log $g$ and microturbulent velocity of the studied 
stars together with [Fe/H] values based on excitation and ionization equilibrium. 
The equivalent widths (EWs) of Fe~I and Fe~II lines and those of other species 
(Mg, Ca, Sc, Ti, Cr, 
Ni, Ba) were measured using DAOSPEC \citep{sp2008} and its up-to-date internal 
list of more than 400 lines, with their respective oscillator 
strengths and excitation potentials. We alternatively checked the EW values of weak lines by 
using the IRAF.SPLOT routine to manually measure the EWs. From the resulting EW list,
we removed those lines with no EW value or with $q$ $>$ 1; $q$ is a quality parameter derived 
from a comparison between the residuals observed in the spectrum in the immediate neighborhood 
of the line and the typical residuals in the spectrum as a whole.

The spectral line analysis to determine the chemical composition of the stars was performed 
with MOOG \citep{sneden1973,sobecketal2011}. MOOG asks mainly for two inputs,
namely, a line data file and a model atmosphere file. The former is mainly the above resulting
DAOSPEC output, with some slight different format arrangement. For the latter, we adopted
the \citet{kurucz2005}'s models generated with the ATLAS code \citep{kurucz1970}.
The models were interpolated for any set of ($T_{eff}$, log $g$, $v_t$, [Fe/H]) using the Python
interpolator \texttt{pyKMOD}\footnote{https://github.com/kolecki4/PyKMOD}. We started
by interpolating the models with the photometric $T_{eff}$ and log $g$ values, and then
by running MOOG to obtain chemical element abundances. We iterated this loop until reaching
convergence at $\Delta$$T_{eff}$ = 5K, $\Delta$log $g$= 0.05 and $\Delta$$v_t$ = 0.05 km/s.
In order to reach excitation and ionization equilibrium, the program simultaneously: 1) varies the values of $T_{eff}$ 
looking for a zero slope in the (FeI,FeII) abundance versus excitation potential relationship; 2) 
 varies $v_t$ looking for a zero slope in the FeI abundance versus wavelength relationship; and 3) 
varies log $g$ looking for similar FeI and FeII abundance values. Since FeI and FeII abundance values changed
as well, each iteration is fed by an interpolated model with the updated overall metallicity.
During the chemical abundance analysis we removed some chemical elements with 
only one EW measure. The resulting stellar atmospheric parameters ($T_{eff}$, log $g$, $v_t$ and [Fe/H])
are listed in Table~\ref{tab1}. The mean and dispersion of [Fe/H] come from the average
of all FeI and FeII abundance values. Table~\ref{tab2} lists the mean and dispersion of the remaining
measured chemical elements, corresponding to the final adopted stellar atmosphere model. 
They correspond to the standard deviation of all the available chemical element abundance values,
which in turn come from the final atmospheric parameters adopted according to the
convergence uncertainties mentioned above.
The difference between the abundance values obtained for star $\#$116
observed in two different nights resulted to be 0.00-0.07 dex, depending on the chemical element.
As for the NGC~6362's red giants we obtained a mean metallicity of [Fe/H] = -1.08$\pm$0.03 dex,
in excellent agreement with \citet{mucciarellietal2016}, so that we did not apply correction
to our metallicities due to systemtic errors.

\begin{table}
\caption{Chemical abundances of the studied stars.}
\label{tab2}
\begin{center}
\begin{tabular}{rccccccc}
\hline\noalign{\smallskip}

ID &  [Mg/Fe]       &  [Ca/Fe]      &  [Sc/Fe]       &    [Ti/Fe]    &  [Cr/Fe]      &  [Ni/Fe]      &   [Ba/Fe] \\       
   &   (dex)        &    (dex)      &    (dex)       &     (dex)     &   (dex)       &   (dex)       &     (dex) \\
\hline\noalign{\smallskip}

44 &   ---          & 0.08$\pm$0.10 & -0.12$\pm$0.12 & 0.07$\pm$0.07 &   ---         & 0.04$\pm$0.12 &   ---     \\ 
48 &   ---          &-0.03$\pm$0.09 &  ---           & 0.05$\pm$0.12 & 0.01$\pm$0.10 &-0.03$\pm$0.13 &   --      -\\ 
49 &   ---          & 0.20$\pm$0.09 &  ---           &    ---        &   ---         & 0.05$\pm$0.13 &  0.03$\pm$0.15\\ 
52 &  0.31$\pm$0.12 & 0.17$\pm$0.12 &   ---          &    ---        &   ---         & 0.08$\pm$0.12 &   ---         \\ 
53 &  ---           &   ---         &  ---           &    ---        &   ---         &       ---     &   ---         \\ 
116&  0.37$\pm$0.15 &  0.02$\pm$0.14& 0.19$\pm$0.15  & 0.20$\pm$0.10 & 0.00$\pm$0.15 & 0.04$\pm$0.15 &   ---         \\ 
118&  0.44$\pm$0.10 & -0.11$\pm$0.13& 0.08$\pm$0.14  & 0.16$\pm$0.14 &   ---         &-0.02$\pm$0.14 & -0.15$\pm$0.16\\ 
119&  ---           &  0.05$\pm$0.12&  ---           & 0.16$\pm$0.13 &   ---         &-0.03$\pm$0.12 & -0.08$\pm$0.16\\ 
126&  ---           &  0.20$\pm$0.14&  ---           & 0.25$\pm$0.15 &   ---         & 0.03$\pm$0.14 &   ---         \\ 
151&   ---          &  0.03$\pm$0.17&  ---           & 0.02$\pm$0.18 &   ---         &-0.09$\pm$0.19 &   ---         \\ 
153&  0.41$\pm$0.12 &  0.32$\pm$0.11&  ---           & 0.06$\pm$0.15 &   ---         & 0.00$\pm$0.12 & -0.03$\pm$0.15\\ 
156&   ---          &  0.13$\pm$0.10& 0.13$\pm$0.13  & 0.18$\pm$0.13 & 0.08$\pm$0.12 & 0.02$\pm$0.13 &   ---         \\ 
157&  0.27$\pm$0.10 &  0.00$\pm$0.12& 0.12$\pm$0.14  & 0.05$\pm$0.14 & 0.09$\pm$0.14 & 0.08$\pm$0.12 &  0.06$\pm$0.16\\ 
168&  0.39$\pm$0.12 &  0.17$\pm$0.12&   ---          & 0.20$\pm$0.12 & 0.05$\pm$0.12 & 0.05$\pm$0.12 & -0.09$\pm$0.14\\ 
173&   ---          &  0.05$\pm$0.10& -0.02$\pm$0.08 & ---           &   ---         & 0.01$\pm$0.08 & -0.04$\pm$0.14\\ 
178&   ---          &  0.09$\pm$0.09&     ---        & 0.05$\pm$0.12 &   ---         & 0.18$\pm$0.06 &  0.02$\pm$0.15\\ 
185&   ---          &  0.24$\pm$0.14&    ---         & 0.32$\pm$0.14 &   ---         &  ---          &   ---         \\ 
186&   ---          &     ---       &    ---         &    ---        &   ---         &  ---          &   ---        \\ 
188&  0.24$\pm$0.08 &  0.10$\pm$0.11&-0.05$\pm$0.11  & 0.00$\pm$0.11 &-0.11$\pm$0.12 & 0.06$\pm$0.12 &   ---        \\ 
191&   ---          &  0.15$\pm$0.16&  ---           & 0.17$\pm$0.17 &   ---         & 0.10$\pm$0.16 &   ---        \\ 
213&  0.16$\pm$0.13 &  0.01$\pm$0.12& 0.13$\pm$0.13  & 0.11$\pm$0.11 &-0.09$\pm$0.13 &-0.03$\pm$0.11 &   ---        \\ 
223&  ---           &  0.16$\pm$0.12& 0.12$\pm$0.12  & 0.03$\pm$0.12 & 0.09$\pm$0.12 & 0.03$\pm$0.11 &   ---        \\ 
224&  0.30$\pm$0.11 &  0.20$\pm$0.14& 0.10$\pm$0.15  & 0.15$\pm$0.15 & 0.05$\pm$0.13 & 0.06$\pm$0.15 & -0.10$\pm$0.17\\
 \noalign{\smallskip}\hline

\end{tabular}
\end{center}
\end{table}

\section{Analysis and discussion}

NGC~6362 has been recently targeted with the aim of accurately estimating its astrophysical properties.
\citet{massarietal2017} obtained ESO@FLAMES.UVES high-resolution spectra for 11 cluster red giant branch stars 
and derived a mean cluster RV and [Fe/H] of -15.03$\pm$2.07 km/s and -1.07$\pm$0.01 dex, respectively.
They also derived mean cluster abundances of 17 chemical elements, among them [Mg/Fe] = 0.54$\pm$0.01 dex, 
[Ca/Fe] = 0.26$\pm$0.02 dex, [Sc/Fe] = 0.18$\pm$0.02 dex, [Ti/Fe] = 0.24$\pm$0.04 dex, [Cr/Fe] = 
-0.05$\pm$0.04 dex, [Ni/Fe] = -0.02$\pm$0.01 dex, and [Ba/Fe] = 0.61$\pm$0.01 dex. We used this cluster
chemical tagging to assess on the origin of a representative sample of stars selected by \citet{kunduetal2019}
as cluster's tidal tail candidates. 

There are several relevant implications from the existence or not of 
NGC~6362's tidal tails, among them whether its tidal tails are kinematic cold or hot, whether NGC~6362 is 
in a regular or chaotic orbital motion; or at what extend criteria of detected globular clusters' tidal 
tails based on kinematic properties are appropriate. Some recent result focused on the analysis of deep images
across an area of  $\sim$ 4 squared degree centered on the cluster \citep{piatti2024}, converged toward a 
relatively smooth stellar density between 1 and $\sim$3.8 cluster Jacobi radii, with a slight difference 
smaller than two times the background stellar density fluctuation between the mean stellar density of
the South-eastern hemisphere and that of the North-western one, with the latter being higher. Moreover, the 
spatial distribution of \citet{kunduetal2019}'s tidal tail stars agrees well not only with the observed 
composite star field distribution, but also with the region least affected by interstellar absorption.
These results would seem to suggest that NGC~6362 would not have clearly detectable tidal tails.

The identification of globular cluster's tidal tail stars has been frequently addressed by looking for stars
that are kinematically consistent with the mean kinematic properties of globular clusters where the stars
formed \citep{sollima2020,xuetal2024}. However, rather than looking for kinematic properties (proper 
motions, RVs) similar to the cluster's ones, it has been suggested in the recent literature to focus on the 
dispersion of the radial and tangential velocities, as well as in the z-component of the angular momentum \citep{malhanetal2021,malhanetal2022}.  This approach  would seem more suitable 
to describe the kinematic properties of tidal tail stars, since stars in order to escape from the cluster 
need to reach velocities different than those of cluster's members. Then, the Milky Way potential imprints 
on them different accelerations, so that mean kinematic properties vary along tidal tail extensions
\citep{piattietal2023,grondinetal2024}.  In order to identify tidal tail stars following this approach, 
the knowledge of the mean path 
of cluster's tidal tail stars in the kinematic space is required, which according to
\citet[][and references therein]{grillmair2025} can be obtained by combining color-magnitude 
diagram and kinematic filtering with orbit integration and predictions based on modeling the stripping 
stars. Alternatively, a kinematic analysis of the sample stars following the approach described in 
\citet{ns2010},  including the use of a Toomre diagram, could also more clearly indicate whether they 
belong to the thin disk, thick disk, or halo populations. Since the main aim of this study is to assess on 
the formation scenario of \citet{kunduetal2019}'s selected stars, we defer such an analysis for a future work.

\begin{figure}
\includegraphics[width=\columnwidth]{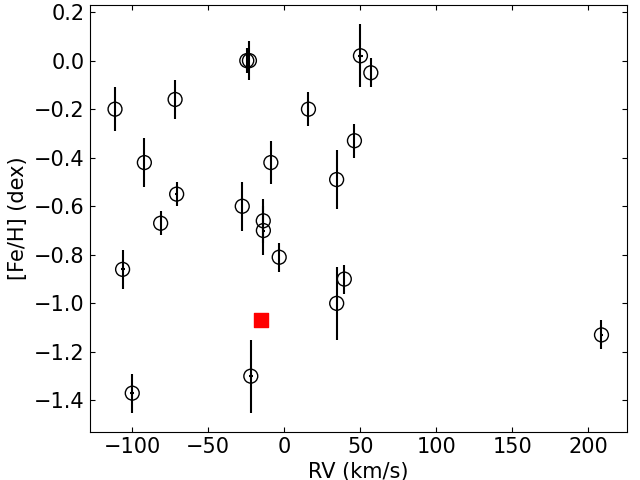}
\caption{[Fe/H] vs RV relationship for the studied stars (open circles, Table~\ref{tab1}) and for
NGC~6362 \citep[red box,][]{massarietal2017}.}
\label{fig2}
\end{figure}

In this context, we probed whether the RVs of \citet{kunduetal2019}'s tidal tail stars are consistent
with the mean cluster's RV, by comparing the resulting RVs (Table~\ref{tab1}) with the mean RV value of 
NGC~6362. We found that two stars ($\#$119 and 126) fall within 1$\sigma$, and other twelve stars
($\#$44, 48, 49, 53, 116, 118, 157, 168, 173, 185, 223 and 224) are within 3$\sigma$,
which represent $\sim$ 9$\%$ and $\sim$ 60$\%$ of the studied star sample, respectively. 
Figure~\ref{fig2} illustrates this finding. We recall that the stars selected by \citet{kunduetal2019}
are within 3$\sigma$ of the mean cluster's proper motion. The above outcome suggests that even in the
most relaxed scenario (RV statistics within 3$\sigma$), a significant percentage of the star sample 
are not coherent with the mean cluster's RV, as it is the case when proper motions are
considered. This discrepancy calls our attention on the possible weakness of selection criteria of tidal 
tail stars based on their kinematic properties. Furthermore, if we restricted the range of RVs to
1$\sigma$ of the mean cluster's RV, the small number of stars that comply with that requirement
tells us that either the 3$\sigma$ sample is contaminated by field stars (because of the large difference
in the number of stars between 1$\sigma$ and 3$\sigma$ samples) or tidal tail stars reach
kinematic properties different than those of the cluster soon after escaping it (because of the spread
in RV of stars located just outside the Jacobi radius). 

In order to find more conclusive evidence on the origin of the studied stars, i.e., to confirm or
dismiss that the studied stars formed in NGC~6362, we analyzed their chemical properties  
\citep[][and references therein]{hankeetal2020}. Unlike kinematic features, the abundance of 
chemical elements remains almost unchanged along the stellar lifetimes. To this respect, 
\citet{marinoetal2019} showed that the metallicity difference between first and second
generation stars in NGC~6362 is $\Delta$[Fe/H] =  0.03 dex. We then compared the resulting
[Fe/H] values with the mean metallicity of NGC~6362 and found that the difference [Fe/H]$_{\rm star}$ - 
[Fe/H]$_{\rm NGC6362}$ is smaller than the respective associated errors added in quadrature
only for stars $\#$52 and 186. Indeed, Figure~\ref{fig2} shows that the spread in metallicty of the 
studied stars is as large as the metallicity dispersion found in the Milky Way disk star field 
population (see below). On the other hand, star $\#$52 has different Mg, Sc, Ti, Cr, and Ba abundances
than NGC~6362's red giants \citep[see][]{massarietal2017}, while star $\#$186 does not share with 
NGC~6362 any of the estimated chemical element abundances. Furthermore, their RVs are remarkably
different from that of the cluster (they are placed just outside the cluster's Jacobi radius), 
even though we assumed that tidal tail star can have RVs different from the mean cluster's RV.
Therefore, it would seem to be unlikely that any of the studied stars have formed in NGC~6362.

The obtained chemical tagging results lead us to conclude that some Milky Way
field stars located along the line-of-sight toward NGC~6362 can have proper motions and positions
in the color-magnitude diagram consistent with the mean cluster's proper motion and cluster's sequences
in the color-magnitude diagram \citep{kunduetal2019}. However, although the latter would seem 
to be a valid selection criterion, the former would not. Indeed, kinematics of highest ranked tidal 
tail stars can show variation of their motions along the tidal tails, as judged by some 
observational \citep{piatti2023b,grillmair2025} and theoretical \citep{grondinetal2024} results,
among others.

\begin{figure*}
\includegraphics[width=\columnwidth]{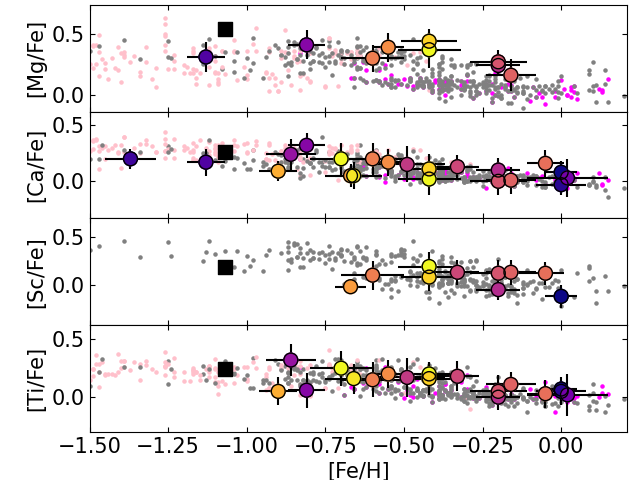}
\includegraphics[width=\columnwidth]{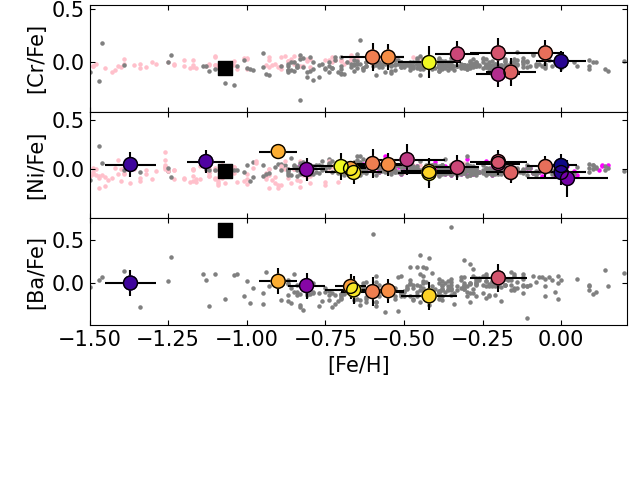}
\caption{[X/Fe] vs [Fe/H] relationships for the studied stars. Small magenta, gray and pink dots represent
Milky Way thin and thick disks and halo field stars, respectively. The large black square represents
NGC~6362.}
\label{fig3}
\end{figure*}

We finally took advantage of the abundances derived for some chemical elements in order to 
further constrain the origin of the studied stars. To this respect, we made use of different 
compilations of Milky Way thick and thin disk and halo field stars, namely: \citet[][thick disk]{reddyetal2003}; 
\citet[][thick and thin disks and halo]{vennetal2004};  \citet[][thick disk]{reddyetal2006}; and 
\citet[][halo]{ns2010}, and built Figure~\ref{fig3}. As can be seen, the [X/Fe] trend with [Fe/H]
has a slope close to zero for Cr and Ni, so that no clear difference comes up for the three
field star populations. In the case of Ba, we did not find any measures for halo stars available.
From Mg, Ca and Ti relationships it is possible to distinguish a somehow bimodal distribution
along the Y-axis for thick disk field stars with some overlap of thin disk field stars
around the most metal-poor peak. Halo stars in general expand the whole measured range
of these chemical element abundances, mainly populating much metal-poor overall metallicities.
From this scenario, we speculate with the possibility that most of the studied stars belong to the
Milky Way disk; some could be part of the thin disk, and a minor percentage might be halo stars.

\section{Conclusions}

We embarked in a spectroscopic tagging analysis of a sample of stars cataloged as
tidal tail star candidates of the Milky Way globular cluster NGC~6362. The
importance of confirming their status as cluster's escaped stars relies on the
consequences for our understanding of globular cluster formation and evolution,
and hence for arriving to a general consensus about the applicable procedures
in searching for globular cluster tidal tail stars. From measures of individual 
radial velocities, metallicities, and abundances of some chemical elements, we
disentangled whether the studied stars formed in NGC~6362 and were later unbound
from the cluster's body due to the interaction with the Milky Way. We concluded that:\\

$\bullet$ Despite the similar values derived of some particular properties for
some stars, none of the studied stars would seem to share the mean cluster chemical element 
abundances within the estimated uncertainties. Therefore, the selected stars, 
which statistically represent nearly 80$\%$ of all the red giant branch tidal
tail candidates, would not seem to be formed in NGC~6362. From the lack of confirmation of
red giant branch tidal tail stars, we speculate that
NGC~6362 does not have detected tidal tails. This conclusion agrees with
the recent outcome by \citet{piatti2024} who showed that the spatial distribution
of the tidal tail star candidates remarkably matches that of the observed 
composite star field distribution, and the regions least affected by interstellar 
absorption.\\

$\bullet$ In agreement with the above result, the measures of individual radial
velocities cover a wide range of values, typical of stars belonging to the composite 
Milky Way field star population. 
Although variation in the radial velocities of tidal tail stars are expected -this 
is not the case, because they have different metallicities-, our present values contrast
with their similar proper motions within 3$\sigma$, and with their location along the 
red giant branch in the cluster color-magnitude diagram \citep{kunduetal2019}. This 
means that kinematics properties would not seem to be as suitable as previously thought for searching
globular cluster tidal tails. Beyond the uncertainties in the {\it Gaia} DR2 data
that could lead to incur in considering some field stars as cluster tidal tail
candidates, the criterion of filtering stars with similar mean globular cluster
kinematics would not seem appropriate. This is because tidal tails already exhibit
a coherent variation of the star motions along them.\\

$\bullet$ The observed stars, distributed along the line-of-sight toward NGC~6362,
would seem to belong mainly to the thick disk, as judged by the derived abundances of some
chemical elements in comparison with previous compilations of Milky Way field
stars. We do not discard the possibility that some of them pertain to the thin disk,
an even a small percentage to the Milky Way halo. The outcome that the studied stars could be
Milky Way field stars agrees well with the cluster having a chaotic orbit \citep{kunduetal2019}, 
which means that its tidal tails were swept while approaching the innermost Milky Way regions.

\normalem
\begin{acknowledgements}
We thank the referee for the thorough reading of the manuscript and
timely suggestions to improve it. 

Based on observations collected at the European Southern Observatory under ESO programme(s) 113.2661.001.

Data for reproducing the figures and analysis in this work will be available upon request
to the author.

\end{acknowledgements}
  

\end{document}